\documentclass[aps,pra,groupedaddress,twocolumn]{revtex4}
\usepackage{amsmath}
\usepackage{amssymb}
\usepackage{epsfig}
\usepackage{graphicx}
\usepackage{times}
\usepackage{amssymb}
\usepackage{hyperref}
\usepackage{textcomp}
\usepackage{xcolor}
\usepackage[T1]{fontenc}

\begin{document}

\title[]{Influence of electron-hole plasma on Rydberg excitons in cuprous oxide}

\author{D. Semkat$^1$, H. Fehske$^1$, and H. Stolz$^2$}
\affiliation{$^1$Institut f\"ur Physik, Ernst-Moritz-Arndt-Universit\"at Greifswald, Felix-Hausdorff-Str.\ 6, 17489 Greifswald, Germany}
\affiliation{$^2$Institut f\"ur Physik, Universit\"at Rostock, Albert-Einstein-Str.\ 23-24, 18059 Rostock, Germany}

\begin{abstract}
We develop a many-body approach to the behavior of exciton bound states and the conduction electron band edge in a surrounding electron-hole plasma with a focus on the absorption spectrum of Rydberg excitons in cuprous oxide. The interplay of band edge and exciton levels is analyzed numerically, whereby the self-consistent solution is compared to the semiclassical Debye approximation. Our results provide criteria which allow to verify or rule out the different band edge models against future experimental data.
\end{abstract}

\maketitle

\section{Introduction}

Since their first experimental confirmation \cite{nature2014}, Rydberg excitons have enjoyed unbowed interest both experimentally and theoretically. A variety of features of such electron-hole bound states with high principal quantum numbers has been investigated since, including high angular momenta \cite{thewes2015}, the behavior in electric and magnetic fields \cite{feldmaierzielinska1617}, the emergence of quantum chaos \cite{assmannostrovskaya2016}, and their interactions \cite{walther2018a} as well as thereby caused giant optical nonlinearities \cite{walther2018b}, to name only a few examples.

Recently it has been shown that a surrounding plasma of free carriers (electrons and holes) has a significant influence on the absorption spectrum, in particular on the position of the band edge, even though its density is very low \cite{heckoetter2018}. Our understanding of magnitude and consequences of this influence is, however, far from being complete. Some features of the absorption lines are very little understood as the temperature dependence of the maximum observable principal quantum number $n_{\rm max}$, to name only a prominent example. Moreover, no shifts of the exciton lines could be found so far which could not be addressed to an insufficient spectral resolution. The latter one is significantly below 1 \textmu eV and, thus, able to resolve the expected line shifts at the relevant densities \cite{heckoetter2018}.
Note that the lack of measured excitonic line shifts has been one of the major objections against the ``plasma hypothesis'' so far. An ongoing, very careful reanalysis of experimental data, however, shows that line shifts exist, but the situation turns out to be quite complex \cite{grossespaper}.
This is one of the main reasons why one should calculate the exciton line and band-edge shifts in the framework of an improved theory selfconsistently.

The behavior of bound states of carriers in a surrounding plasma has been the subject of well elaborated theories since decades where both atomic plasmas and electron-hole plasmas in semiconductors have been investigated \cite{rotesbuch,gruenesbuch,zimmermann88,kremp2005}. Naturally, the principle effects have been studied at the ground ($1S$-)state or at the lower excited states. Here, we aim at much higher excited states with principal quantum numbers up to $n>10$. The peculiarity of these states is the low binding energy (which scales, as is known, like $n^{-2}$) so that they are situated energetically just below the band edge.
These Ryd\-berg states, therefore, are expected (and, qualitatively, have been shown \cite{heckoetter2018}) to be extremely sensitive already against very thin electron-hole plasmas.

While the line shifts are known to be small over a broad range of densities and well approximated by first order perturbation theory, the band edge shifts much stronger. Therefore it has to be determined by self-consistently solving an intergral equation, if analyzed on the same theoretical level (dynamically screened approximation) as the line shifts. For the solution of that equation, several approximations are used in the literature \cite{seidel95}.

In the following, we outline a consistent many-body theoretical approach to band edge and excitonic line shifts. We discuss the obtained numerical results with the focus on the interplay of band edge and exciton states and a comparison of different approximations for the band edge. The conclusions drawn from that are presented as predictions which have to be verified by comparison to data from recent and ongoing experiments.

\section{Theoretical approach}

\subsection{Band edge}

The band edge, i.e., the lower edge of the conduction band, is given by the lowest possible energy for scattering states of electrons and holes. Therefore, the many-body induced shift of that quantity should follow from solving the effective wave equation for electron-hole pairs (Bethe--Salpeter equation) by inserting the asymptotic wave function of scattering states \cite{seidel95}, $\psi(\mathbf{k}) = (2\pi)^3\delta(\mathbf{k})$. One obtains the intuitive result \cite{footnote0} $E_{\rm be}=E_{\rm e}(\mathbf{k}=\mathbf{0})+E_{\rm h}(\mathbf{k}=\mathbf{0})$. The dispersions of electrons and holes read in (self-consistent) quasiparticle approximation
\begin{equation}\label{E1qp}
E_a(\mathbf{k})=\frac{\hbar^2k^2}{2m_a}+\mathrm{Re}\Sigma(\mathbf{k},\omega=E_a(\mathbf{k})/\hbar)\,,\quad a=\mathrm{e,h}\,.
\end{equation}
Using the self-energy in dynamically screened approximation (so-called $V^{\rm s}$ approximation) \cite{seidel95}, we get for the band edge
\begin{eqnarray}\label{bandkante}
\hspace*{-2ex}E_{\rm be}&=&\Sigma_{\rm e}^{\rm HF}(0)+\Sigma_{\rm h}^{\rm HF}(0)\nonumber\\
&&+\mathcal{P}\int\frac{\mathrm{d}^3k}{(2\pi)^3}\,V_{\rm eh}(k)\int\frac{\mathrm{d}(\hbar\omega)}{\pi}\,\mathrm{Im}\,\varepsilon^{-1}(k,\omega)\nonumber\\
&&\times\left[\frac{n_{\rm B}(\omega)+1}{E_{\rm be}-\hbar\omega-E_{\rm e}(\mathbf{k})-E_{\rm h}(\mathbf{0})}\right.\nonumber\\
&&\left.+
\frac{n_{\rm B}(\omega)+1}{E_{\rm be}-\hbar\omega-E_{\rm h}(\mathbf{k})-E_{\rm e}(\mathbf{0})}\right]
\end{eqnarray}
with the Hartree--Fock self-energy
\begin{equation}\label{shf}
\Sigma_a^{\rm HF}(\mathbf{k})=-\int\frac{\mathrm{d}^3k'}{(2\pi)^3}\,V_{aa}(|\mathbf{k}-\mathbf{k}'|)f_a(\mathbf{k}')
\end{equation}
(note that the Hartree term vanishes due to electroneutrality), the Coulomb potential $V_{ab}(k)=e_ae_b/(\varepsilon_0\varepsilon_{\rm b}k^2)$ (with $a,b=\mathrm{e,h}$ and $\varepsilon_{\rm b}$ being the background dielectric constant), the dielectric function $\varepsilon(\mathbf{k},\omega)$, the Bose distribution\linebreak $n_{\rm B}(\omega)=1/\left(\mathrm{exp}\left[\hbar\omega/(k_{\rm B}T)\right]-1\right)$, and the Fermi distribution $f_a(\mathbf{k})=1/\left(\mathrm{exp}\left[\left(E_a(\mathbf{k})-\mu_a\right)/(k_{\rm B}T)\right]+1\right)$.

The frequency integral can be evaluated \cite{footnote1} leading to
\begin{eqnarray}\label{bandkante1}
E_{\rm be}&=&\Sigma_{\rm e}^{\rm HF}(0)+\Sigma_{\rm h}^{\rm HF}(0)
-\frac{e^2}{\varepsilon_0\varepsilon_{\rm b}}\frac{1}{(2\pi)^2}\int\limits_0^{\infty}\mathrm{d}k
\\
&&\hspace*{-9ex}\times\Bigg\{
[1+n_{\rm B}(\omega_{\rm e}(k))]\left[\mathrm{Re}\,\varepsilon^{-1}(k,\omega_{\rm e}(k))-1\right]\nonumber\\
&&\hspace*{-7ex}-\frac{k_{\rm B}T}{\hbar\omega_{\rm e}(k)}\left[\varepsilon^{-1}(k,0)-1\right]\nonumber\\
&&\hspace*{-7ex}-\frac{4\hbar\omega_{\rm e}(k)}{k_{\rm B}T}\sum\limits_{j=1}^{\infty}\frac{\varepsilon^{-1}(k,\mathrm{i}2\pi jk_{\rm B}T/\hbar)-1}
{\left(\frac{\hbar\omega_{\rm e}(k)}{k_{\rm B}T}\right)^2+(2\pi j)^2}
+\left\{\omega_{\rm e}\leftrightarrow\omega_{\rm h}\right\}\Bigg\}\nonumber
\end{eqnarray}
with
\begin{equation}\label{omegaeh}
\hbar\omega_{\rm e/h}(k)=E_{\rm be}-E_{\rm e/h}(k)-E_{\rm h/e}(0)\,.
\end{equation}

In principle, Eq.\ (\ref{bandkante1}) has to be solved along with the self-consistent solution of Eq.\ (\ref{E1qp}) and the corresponding equation for $\Sigma$. There are two more or less obvious possibilities to simplify the problem by breaking up the full self-consistency: (i) If one approximates the quasiparticle energies in the integrand of Eq.\ (\ref{bandkante}) by free-particle energies and correspondingly the band edge by its unperturbed value of 0, i.e., approximating
\begin{equation}\label{bandkanters}
\hbar\omega_{\rm e/h}(k)=-\frac{\hbar^2k^2}{2m_{\rm e/h}}\,,
\end{equation}
one obtains a closed expression $E_{\rm be}^{(0)}$. Exactly the same expression is obtained by applying the rigid-shift approximation, i.e., replacing the self-energy by a $k$-independent shift \cite{5maenner,kremp2005}. Then $E_{\rm be}$ in the denominators is given by the sum of electron and hole rigid shifts which cancel the corresponding terms in the dispersions.
(ii) To go one step beyond that approximation, in Ref.\ \cite{seidel95} a slightly different approach has been proposed, i.e., using the same approximations for the quasiparticle energies but leaving the band edge in $\omega_{\rm e/h}$ unchanged,
\begin{equation}\label{bandkanteiter}
\hbar\omega_{\rm e/h}(k)=E_{\rm be}-\frac{\hbar^2k^2}{2m_{\rm e/h}}\,,
\end{equation}
leading to a self-consistent equation for $E_{\rm be}$ that can be solved iteratively ($E_{\rm be}^{\rm iter}$).

It is important to note, however, that one has to be very careful to keep the same level of approximation both for band edge and quasiparticle energies. This condition is violated in (\ref{bandkanteiter}) since $E_{\rm be}$ consists of just those self-energies which have been omitted in deriving Eq.\ (\ref{bandkanteiter}) from (\ref{omegaeh}). Therefore, we modify the idea from Ref.\ \cite{seidel95} by stating that electrons and holes would contribute equally to the band edge (\ref{bandkante}) if they had equal masses. Otherwise it holds at least approximately that $E_{\rm e/h}(\mathbf{0})\approx\frac{1}{2}E_{\rm be}$. Inserting this into $\omega_{\rm e/h}$ leads to a slightly modified expression (note the prefactor 1/2),
\begin{equation}\label{bandkanteiter1}
\hbar\omega_{\rm e/h}(k)=\frac{1}{2}E_{\rm be}-\frac{\hbar^2k^2}{2m_{\rm e/h}}\,,
\end{equation}
leading again to a self-consistent equation for $E_{\rm be}$ the solution of which will be denoted by $E_{\rm be}^{\rm mod}$ lateron.

The approximations given above can be compared to the semiclassical Debye shift,
\begin{eqnarray}\label{debye}
E_{\rm be}^{\rm Debye}&=&-\frac{1}{4\pi\varepsilon_0\varepsilon_{\rm b}}\kappa e^2
\end{eqnarray}
with $\kappa$ being the inverse screening length,
\begin{eqnarray}\label{kappa}
\kappa^2=\frac{e^2}{\varepsilon_0\varepsilon_{\rm b}}\sum\limits_{a=\mathrm{e,h}}\frac{\partial n_a}{\partial\mu_a}
=\frac{e^2}{\varepsilon_0\varepsilon_{\rm b}}\sum\limits_{a=\mathrm{e,h}}\frac{n_a}{k_{\rm B}T_a}\,,
\end{eqnarray}
the latter relation holding in the nondegenerate (Debye) case.

\subsection{Exciton lines}

The exciton energies as well as their wave functions are the bound state solutions of the effective wave equation (homogeneous Bethe--Salpeter equation, effective two-particle Schr\"odinger equation) \cite{5maenner,gruenesbuch}. Dividing the two-particle (electron-hole) Hamiltonian into the part describing the isolated two-particle pair and a part $\mathcal{H}'$ comprising all contributions of many-body effects, the latter one reads for vanishing center-of-mass momentum of the exciton
\begin{eqnarray}\label{hamiltonian}
&&\mathcal{H}'\psi(\mathbf{k},\omega)=-\int\frac{\mathrm{d}^3k'}{(2\pi)^3}\nonumber\\
&&\times\left\{
V_{\rm eh}(|\mathbf{k}-\mathbf{k}'|)\left[f_{\rm e}(\mathbf{k})+f_{\rm h}(-\mathbf{k})\right]
\psi(\mathbf{k}',\omega)\right.\nonumber\\
&&\vspace*{10ex}
-V_{\rm eh}(|\mathbf{k}-\mathbf{k}'|)\left[f_{\rm e}(\mathbf{k}')+f_{\rm h}(-\mathbf{k}')\right]\psi(\mathbf{k},\omega)\nonumber\\
&&\vspace*{10ex}
\left.+\Delta V_{\rm eh}^{\rm eff}(\mathbf{k},\mathbf{k}',\omega)\left[\psi(\mathbf{k}',\omega)-\psi(\mathbf{k},\omega)\right]
\right\}
\end{eqnarray}
with $\psi(\mathbf{k},\omega)$ being the excitonic wave function.
The effective potential $\Delta V_{\rm eh}^{\rm eff}$ is given by
\begin{eqnarray}\label{veff}
&&\Delta V_{\rm eh}^{\rm eff}(\mathbf{k},\mathbf{k}',\omega)=V_{\rm eh}(|\mathbf{k}-\mathbf{k}'|)\\
&&\times\int\limits_{-\infty}^{\infty}\frac{\mathrm{d}\bar{\omega}}{\pi}\,
\mathrm{Im}\,\varepsilon^{-1}(\mathbf{k}-\mathbf{k}',\bar{\omega}+\mathrm{i}\epsilon)\nonumber\\
&&\times\left\{
\frac{n_{\rm B}(\bar{\omega})+1}{\hbar\omega+\mathrm{i}\epsilon-\hbar\bar{\omega}-E_{\rm e}(\mathbf{k}')-E_{\rm h}(-\mathbf{k})}\right.\nonumber\\
&&+\left.
\frac{n_{\rm B}(\bar{\omega})+1}{\hbar\omega+\mathrm{i}\epsilon-\hbar\bar{\omega}-E_{\rm e}(\mathbf{k})-E_{\rm h}(-\mathbf{k}')}
\right\}\,.\nonumber
\end{eqnarray}

One can identify four fundamental many-body effects in the electron-hole Hamiltonian (\ref{hamiltonian}), (i) phase space occupation or Pauli blocking, respectively (second line), (ii) exchange (Fock) self-energy (third line), (iii) dynamical self-energy correction (fourth line, first term), and (iv) dynamically screened effective potential (fourth line, second term) \cite{5maenner,gruenesbuch}.
Using perturbation theory, the first-order correction to the energy of an excitonic state is then given by the expectation value of $\mathcal{H}'$ with the unperturbed two-particle wave functions,
\begin{eqnarray}\label{DeltaE}
\Delta E_{nl}^{(1)}&=&\int\mathrm{d}^3k\,\psi^*(\mathbf{k},\omega)\mathcal{H}'\psi(\mathbf{k},\omega)\big|_{\hbar\omega=E_{nl}^{(0)}}\nonumber\\
&=&\Delta E_{nl}^{\rm PB}+\Delta E_{nl}^{\rm F}+\Delta E_{nl}^{\rm DS}+\Delta E_{nl}^{\rm DP}\,.
\end{eqnarray}
For the evaluation of the four expectation values, we factorize the wave function in the usual manner into radial part and spherical harmonics,
$\psi(\mathbf{k})=\phi_{nl}(k)Y_{lm}(\vartheta,\varphi)$.
Summarizing the ``static'' and the ``dynamic'' contributions, respectively, one obtains for the Pauli blocking and Fock contributions
\begin{eqnarray}\label{deltaepbf}
&&\Delta E_{nl}^{\rm PBF}=\Delta E_{nl}^{\rm PB}+\Delta E_{nl}^{\rm F}\\
&&=\frac{e^2}{\varepsilon_0\varepsilon_{\rm b}}\frac{1}{(2\pi)^2}\int\limits_0^{\infty}\mathrm{d}k\int\limits_0^{\infty}\mathrm{d}k'\,kk'\left[f_{\rm e}(k)+f_{\rm h}(k)\right]\phi_{nl}(k')\nonumber\\
&&\times\left\{ \phi_{nl}^*(k)Q_l\left(\frac{k^2+k'^2}{2kk'}\right)-\phi_{nl}^*(k')Q_0\left(\frac{k^2+k'^2}{2kk'}\right)\right\}\,,\nonumber
\end{eqnarray}
where $P_l$ is the Legendre polynomial and $Q_l$ is the Legendre function of the second kind,
$Q_l(z)=\frac{1}{2}\int_{-1}^1\mathrm{d}t\,P_l(t)/(z-t)$,
and for the contributions arising from dynamical self-energy and effective potential \cite{footnote1}
\begin{eqnarray}\label{deltaedsp}
&&\Delta E_{nl}^{\rm DSP}=\Delta E_{nl}^{\rm DS}+\Delta E_{nl}^{\rm DP}\\
&&=
-\frac{e^2}{\varepsilon_0\varepsilon_{\rm b}}\frac{1}{(2\pi)^2}\int\limits_0^{\infty}\mathrm{d}k\int\limits_0^{\infty}\mathrm{d}k'\,kk'\,
\int\limits_{-1}^1\mathrm{d}t\,\frac{1}{\frac{k^2+k'^2}{2kk'}-t}\nonumber\\
&&\times\left\{ \phi_{nl}^*(k)\phi_{nl}(k')P_l(t)-\frac{1}{2}\left[|\phi_{nl}(k)|^2+|\phi_{nl}(k')|^2\right]\right\}\nonumber\\
&&\times\Bigg\{
[1+n_{\rm B}(\omega_0(k,k'))]\nonumber\\
&&\vspace*{2ex}\times\left[\mathrm{Re}\,\varepsilon^{-1}(\sqrt{k^2+k'^2-2kk't},\omega_0(k,k'))-1\right]\nonumber\\
&&-\frac{k_{\rm B}T}{\hbar\omega_0(k,k')}
\left[\varepsilon^{-1}(\sqrt{k^2+k'^2-2kk't},0)-1\right]\nonumber\\
&&-\frac{4\hbar\omega_0(k,k')}{k_{\rm B}T}\sum\limits_{j=1}^{\infty}\frac{\varepsilon^{-1}(k,\mathrm{i}2\pi jk_{\rm B}T/\hbar)-1}
{\left(\frac{\hbar\omega_0(k,k')}{k_{\rm B}T}\right)^2+(2\pi j)^2}\Bigg\}\nonumber
\end{eqnarray}
with
\begin{eqnarray}
\hbar\omega_0(k,k')=E_n-\frac{\hbar^2k^2}{2m_{\rm e}}-\frac{\hbar^2k'^2}{2m_{\rm h}}\,.
\end{eqnarray}

Hereinafter, the dielectric function $\varepsilon(\mathbf{k},\omega)$ will be used in the nondegenerate limit of the ``random phase approximation'' (RPA). It reads in excitonic (Heaviside) units \cite{fehr94,footnote2}
\begin{eqnarray}\label{reepsilonrpa}
\mathrm{Re}\,\varepsilon(X,Y)&=&1+\frac{K^2}{4X^2}\sum\limits_a\sum\limits_{j=0}^1\,(-1)^j\,A_a(X,Y)\nonumber\\
&&\times_1F_1\left[1,\frac{3}{2};-\frac{A_a^2(X,Y)K^2X^2}{4\delta_a}\right]\,,\\\label{imepsilonrpa}
\mathrm{Im}\,\varepsilon(X,Y)&=&-\frac{\sqrt{\pi}}{4}\frac{K}{X^3}\sum\limits_a\sum\limits_{j=0}^1\,\sqrt{\delta_a}(-1)^j\nonumber\\
&&\times\mathrm{e}^{-\frac{A_a^2(X,Y)K^2X^2}{4\delta_a}}
\end{eqnarray}
with
$A_a(X,Y)=\delta_aY/X^2+(-1)^j$, $K^2=\hbar\omega_{\rm pl}^{\rm e}/(k_{\rm B}T)$, $X^2=\hbar^2q^2/(2m_{\rm e}\,\hbar\omega_{\rm pl}^{\rm e})$, $Y=\omega/\omega_{\rm pl}^{\rm e}$, $\delta_a=m_a/m_{\rm e}$, $\omega_{\rm pl}^{\rm e}$ is the electron plasma frequency, $(\omega_{\rm pl}^{\rm e})^2=n_{\rm e}e^2/(\varepsilon_0\varepsilon_{\rm b}m_{\rm e})$,
and $_1F_1$ is the confluent hypergeometric function.

In the static limit, the Hamiltonian (\ref{hamiltonian}) gets a simpler form \cite{seidel95,arndt96}. Applying in addition the classical (Debye) limit, the shift (\ref{deltaedsp}) turns into
\begin{eqnarray}\label{deltaedebye}
&&\Delta E_{nl}^{\rm Debye}=
-\frac{e^2}{\varepsilon_0\varepsilon_{\rm b}}\frac{1}{(2\pi)^2}\int\limits_0^{\infty}\mathrm{d}k\int\limits_0^{\infty}\mathrm{d}k'\,kk'\\
&&\times\int\limits_{-1}^1\mathrm{d}t\,\left\{\frac{1}{\frac{k^2+k'^2+\kappa^2}{2kk'}-t}-\frac{1}{\frac{k^2+k'^2}{2kk'}-t}\right\}\nonumber\\
&&\times\left\{ \phi_{nl}^*(k)\phi_{nl}(k')P_l(t)-\frac{1}{2}\left[|\phi_{nl}(k)|^2+|\phi_{nl}(k')|^2\right]\right\}\,.\nonumber
\end{eqnarray}
Note that the part of the integrand arising from the second term in the last line (with prefactor 1/2) can be integrated out providing the well-known Debye shift $-\kappa e^2/(4\pi\varepsilon_0\varepsilon_{\rm b})$. For numerical reasons, it is advisible to perform the numerical integration on the whole integrand.

The determination of the excitonic lineshifts by perturbation theory can be justified, in principle, only by comparison to the exact solution of the problem (solution of the Bethe-Salpeter equation for the electron-hole bound states). That has been done in plasma physics \cite{fehr94} showing that first order perturbation theory gives very good results compared to the exact solution even close to the Mott transition.

While Eq.\ (\ref{hamiltonian}) represents only the Hermitian part of the plasma Hamiltonian with real eigenvalues giving rise to the shift of the exciton line, the non-Hermitian part yields the linewidth (and, therefore, the plasma-induced lifetime of the state). The resulting linewidths can be shown to be small (which justifies the two-particle description of the electron-hole pairs) and are not considered explicitly here. We refer to the detailed discussion of this matter in Refs.\ \cite{seidel95,arndt96}. The actual excitonic linewidths can be assigned to other physical processes, i.e., scattering with acoustic and optical phonons \cite{stolz2018}.

\section{Numerical results}

\begin{figure}[tbh]%
\begin{center}
\includegraphics[width=0.9\columnwidth]{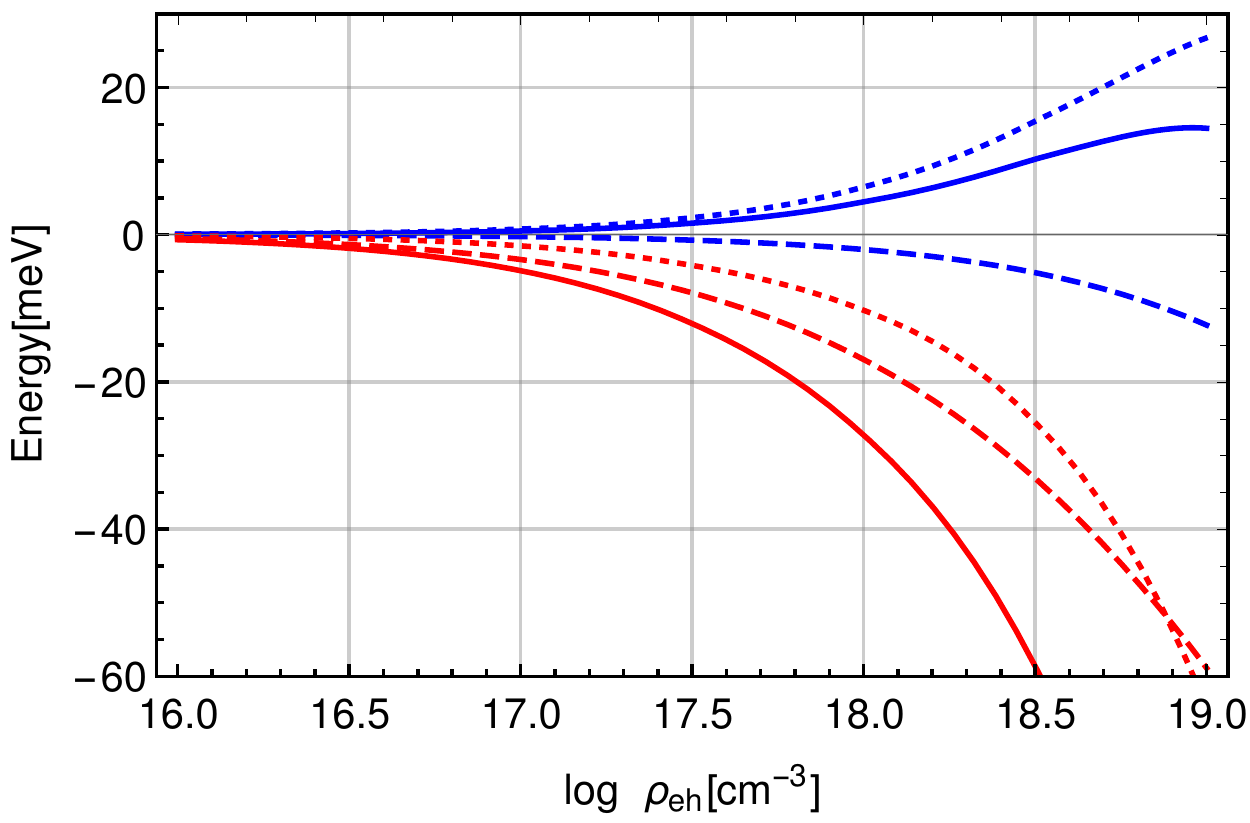}
\end{center}
\caption{Contributions to the excitonic line shift for the $1S$ (blue) and $2P$ (red) excitons at $T=10$ K \textit{vs.}\ density $\rho_{\rm eh}$. Sum of Fock and Pauli blocking contributions $\Delta E_{nl}^{\rm PBF}$ [Eq.\ (\ref{deltaepbf})] (dotted) and sum of the dynamic contributions $\Delta E_{nl}^{\rm DSP}$ [Eq.\ (\ref{deltaedsp})] (dashed). The solid lines denote the total shift $\Delta E_{nl}^{\rm PBF}+\Delta E_{nl}^{\rm DSP}$, cf.\ Eq.\ (\ref{DeltaE}).
}
\label{fig:edps-epdf-1s2p}
\end{figure}
Before analyzing the interplay of band edge and exciton levels, we take a closer look onto the different contributions to the excitonic line shifts. Figure \ref{fig:edps-epdf-1s2p} shows, for $1S$- and $2P$-excitons, the contributions $\Delta E_{nl}^{\rm PBF}$, Eq.\ (\ref{deltaepbf}) and $\Delta E_{nl}^{\rm DSP}$, Eq.\ (\ref{deltaedsp}) in dependence on plasma density $\rho_{\rm eh}$ for a temperature of $T=10$ K. For the $2P$-exciton, both contributions to the shift are negative and of the same order of magnitude. For the $1S$-state, however, $\Delta E_{nl}^{\rm PBF}$ is positive and, over a broad density range, larger than $\Delta E_{nl}^{\rm DSP}$, causing the total shift to be positive.
This behavior is well known from hydrogen as well as electron-hole plasmas, see, e.g., \cite{seidel95,arndt96}. At least in cuprous oxide, it seems to be restricted to the $1S$-line. Note that $\Delta E_{nl}^{\rm PBF}$ decreases for higher temperatures and the total shift becomes negative again.

\begin{figure}[tbh]%
\begin{center}
\includegraphics[width=0.9\columnwidth]{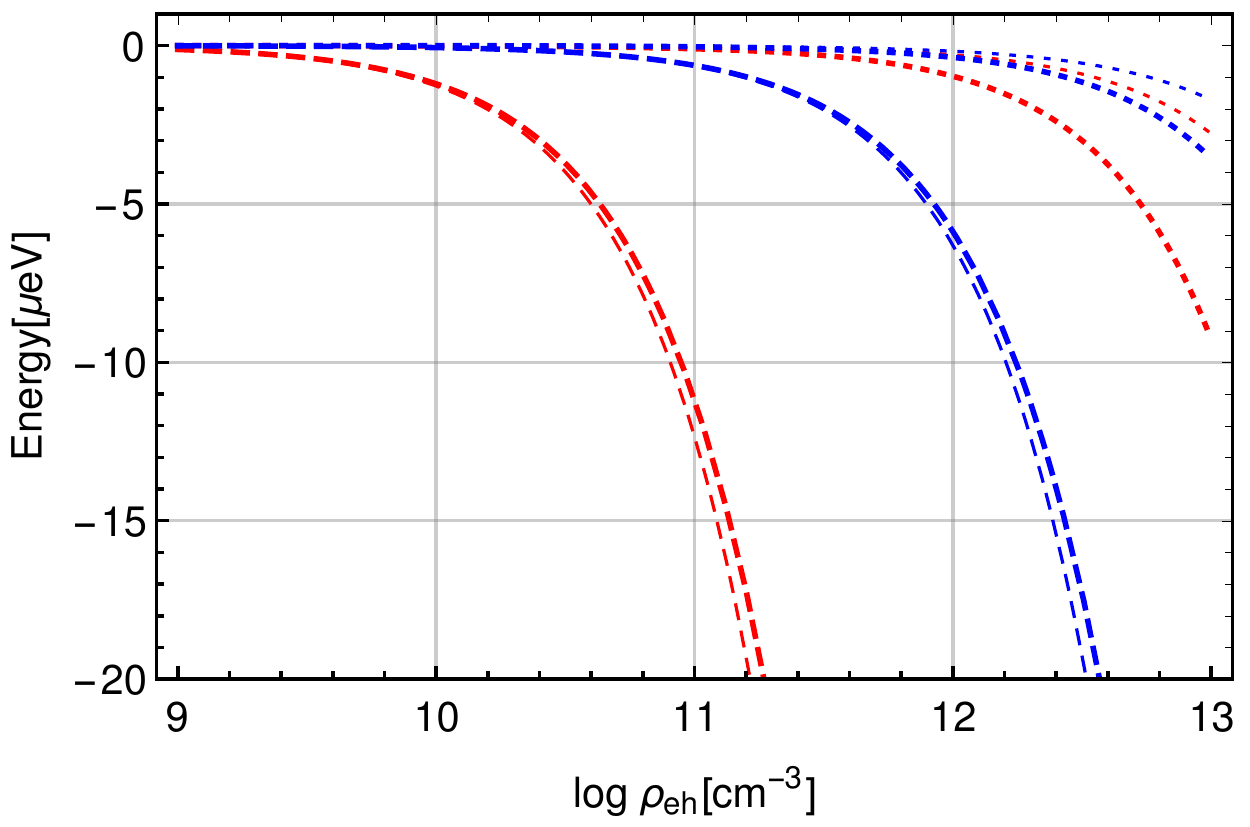}
\end{center}
\caption{Same representation as in Fig.\ \ref{fig:edps-epdf-1s2p}, but for the $n=5$ (blue) and $n=10$ (red) $P$-excitons at $T=2$ K (thick lines) and $T=10$ K (thin lines) \textit{vs.}\ density $\rho_{\rm eh}$. For obvious reasons, the total shift $\Delta E_{nl}^{\rm PBF}+\Delta E_{nl}^{\rm DSP}$ is not shown.
}
\label{fig:edps-epdf-n5n10}
\end{figure}
Going from the $2P$-level to higher states, the balance between both contributions to the shift changes considerably. Figure \ref{fig:edps-epdf-n5n10} shows the contributions for $P$-excitons of two different principal quantum numbers $n$ (5 and 10) and two different plasma temperatures $T$ (2 K and 10 K) in dependence on plasma density $\rho_{\rm eh}$. Since only quite low densities are experimentally relevant \cite{heckoetter2018}, the Fock and Pauli blocking contributions are, both for $n=5$ and $n=10$, very small compared to those arising from dynamical self-energy and dynamically screened potential, even for the lower temperature of 2 K. While $\Delta E_{nl}^{\rm PBF}$ decreases with increasing temperature, $\Delta E_{nl}^{\rm DSP}$ increases slightly.

\begin{figure}[tbh]%
\begin{center}
\includegraphics[width=0.9\columnwidth]{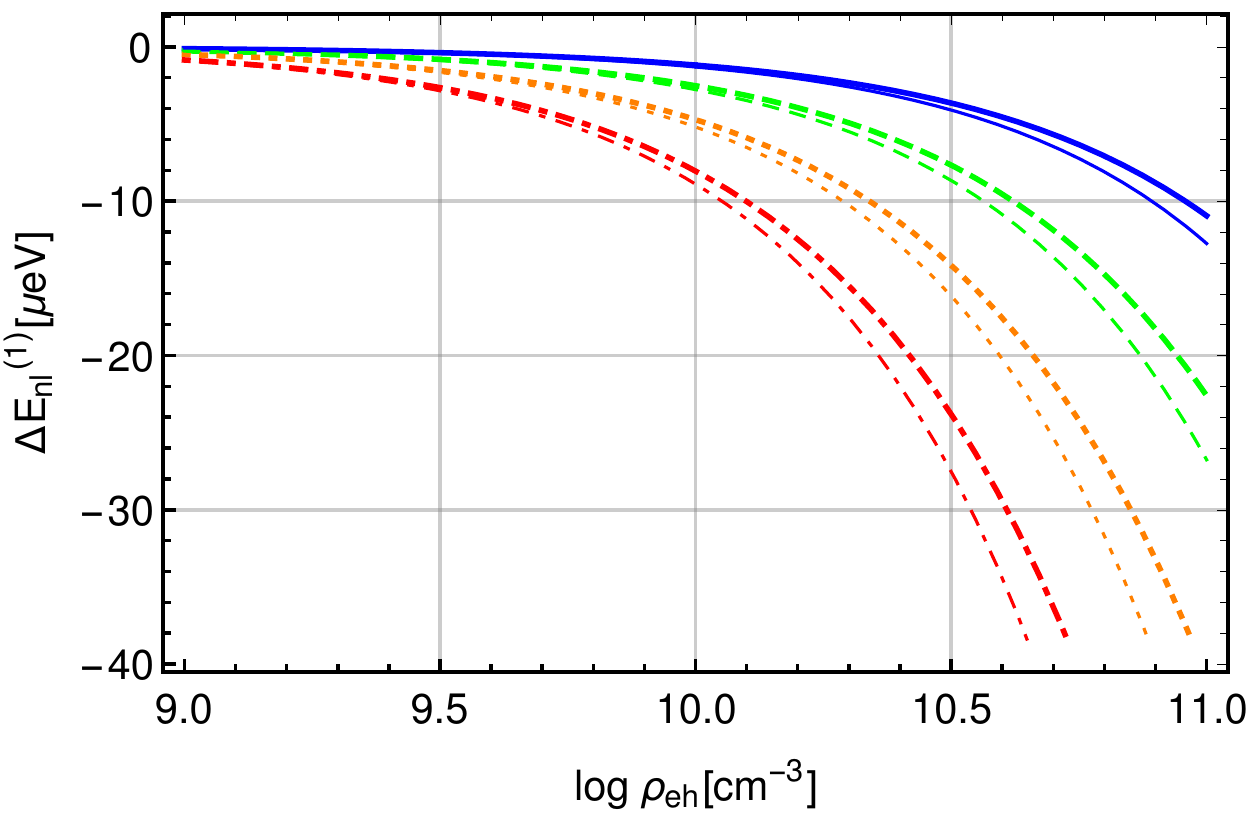}
\end{center}
\caption{Total excitonic line shift $\Delta E_{nl}^{(1)}$, Eq.\ (\ref{DeltaE}), for the $n=10$ (solid blue line), $n=12$ (green dashed), $n=14$ (orange dotted), and $n=16$ (red dash-dotted) $P$-excitons at $T=2$ K (thick lines) and $T=18$ K (thin lines) \textit{vs.}\ density $\rho_{\rm eh}$.
}
\label{fig:deltae10-16}
\end{figure}
Figure \ref{fig:deltae10-16} shows the line shifts for several principal quantum numbers in dependence on plasma density for two temperatures. In the depicted density region relevant for the experiments, the temperature dependence of the shifts is obviously quite weak. Strikingly, the shifts are larger for higher $n$ while the total energies, of course, decrease with $n^{-2}$. We will come back to this point when discussing Fig.\ \ref{fig:mott121416}. A quantitative analysis of the shifts reveals a scaling law the leading term of which is given by $\Delta E_{nl}=c\,n^4\rho_{\rm eh}$ with $c=0.0133$ \textmu eV \textmu m$^3$ for low densities $\rho_{\rm eh}$ and higher principal quantum numbers $n$. The asymptotically linear density dependence agrees with the result obtained by Seidel \textit{et al.} \cite{seidel95}. Since the bare Ryd\-berg energies scale approximately like $n^{-2}$, we find for the relative energy shifts $\Delta E_{nl}/E_{nl}\propto n^6\rho_{\rm eh}$, i.e., the higher the principal quantum number is, the more sensitive is the respective exciton state to the influence of the electron-hole plasma.
\begin{figure}[tbh]%
\begin{center}
\includegraphics[width=0.9\columnwidth]{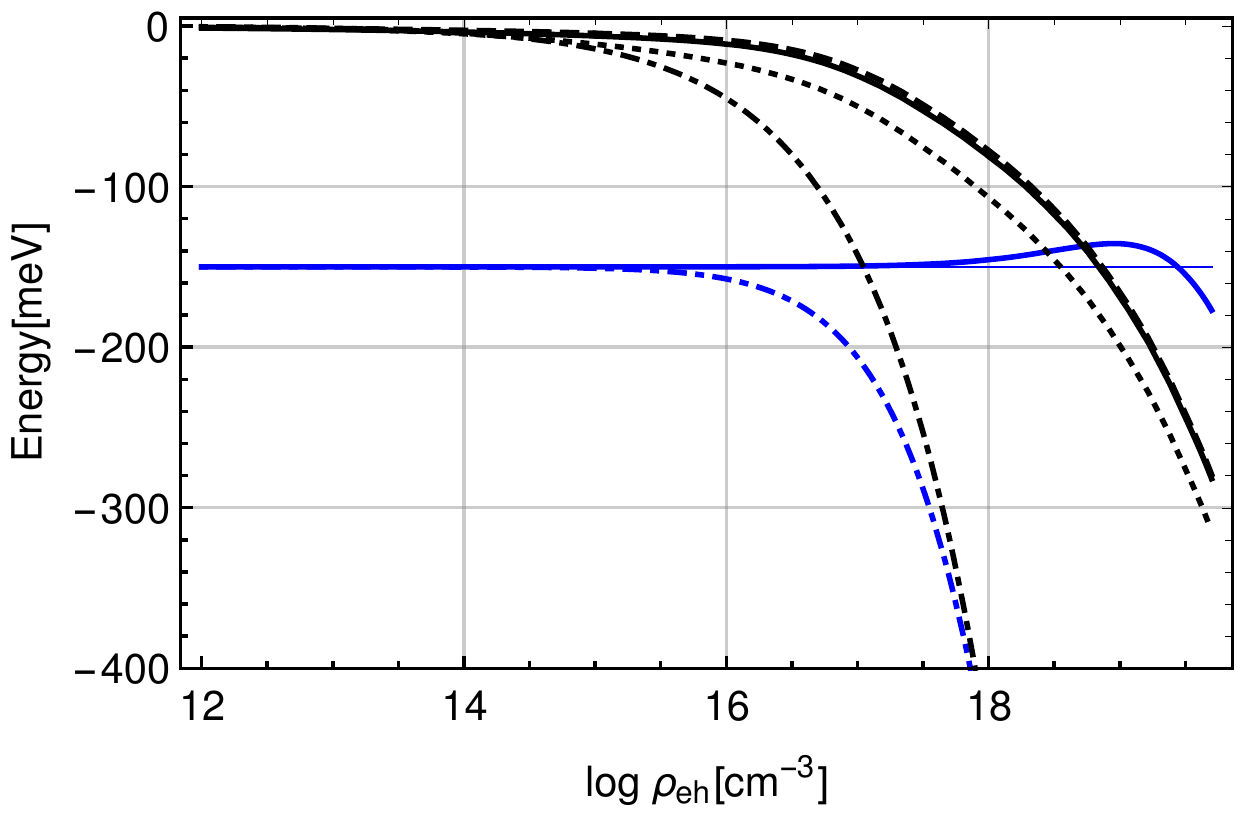}\\
\vspace*{0.2cm}

\includegraphics[width=0.9\columnwidth]{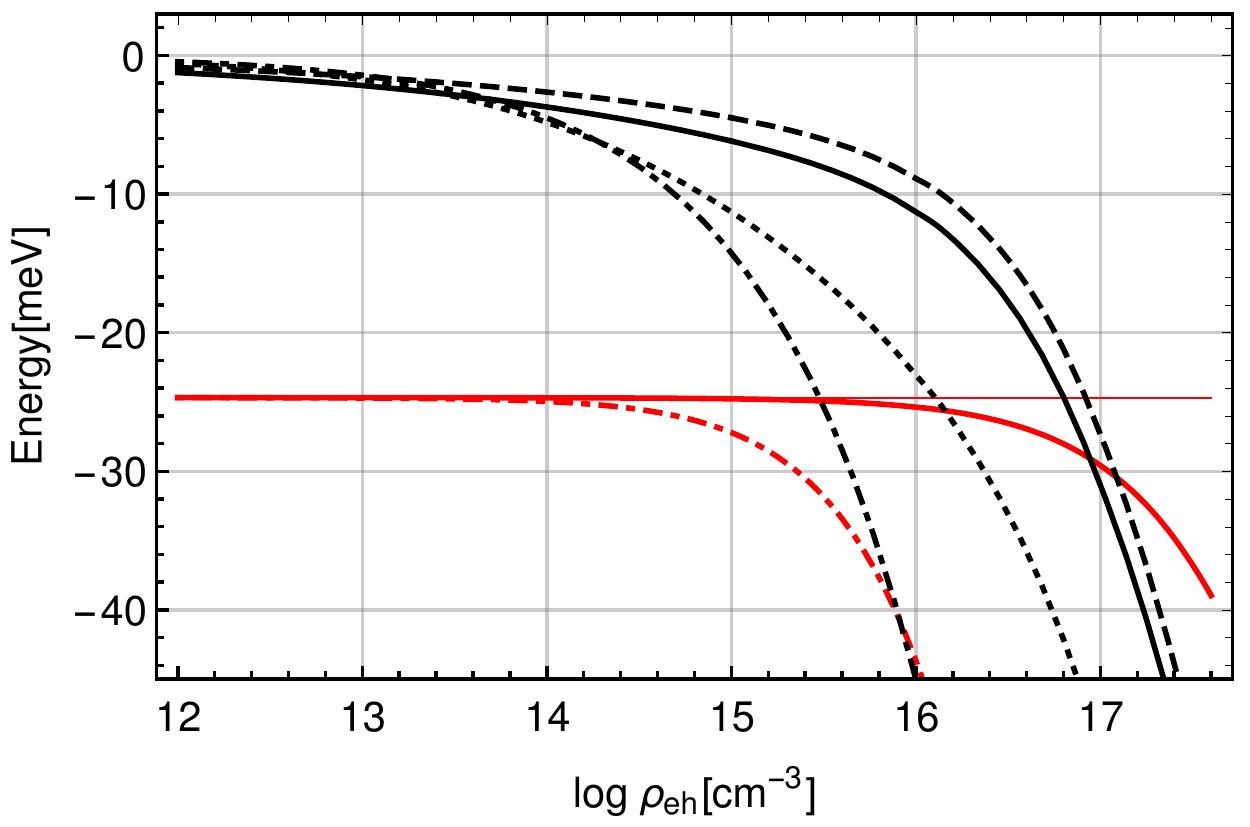}
\end{center}
\caption{Band edge (black lines) and $n=1$ $S$-exciton level (blue lines; upper panel) and $n=2$ $P$-exciton level (red lines; lower panel) \textit{vs.}\ density for a carrier temperature of $T=10$ K. Approximations for the band edge: $E_{\rm be}^{\rm mod}$ (Eq.\ (\ref{bandkante}) with (\ref{bandkanteiter1}); solid line), $E_{\rm be}^{\rm iter}$ (Eq.\ (\ref{bandkante}) with (\ref{bandkanteiter}); dashed), $E_{\rm be}^{(0)}$ (Eq.\ (\ref{bandkante}) with (\ref{bandkanters}); dotted), and $E_{\rm be}^{\rm Debye}$ (Eq.\ (\ref{debye}); dash-dotted). Thick (thin) colored lines denote the shifted (unperturbed) exciton levels, whereupon the shift is given by $\Delta E_{nl}^{(1)}$ (Eq.\ (\ref{DeltaE}); solid) and $\Delta E_{nl}^{\rm Debye}$ (Eq.\ (\ref{deltaedebye}); dash-dotted).
}
\label{fig:mott1s2p}
\end{figure}

Let us now turn to the interplay of band edge and exciton levels. To start with, we look again at the $1S$-level.
Figure \ref{fig:mott1s2p} (upper panel) shows band edge and $1S$-exciton energy in dependence on the electron-hole plasma density. The principle behavior is well known \cite{rotesbuch,5maenner,haug78}: The band edge (black lines) shifts down with increasing density while the exciton level (red lines) remains constant at first due to a wide compensation of many-body effects. At higher densities, the compensation becomes incomplete and the level shifts, too (down or up depends, as discussed above, on quantum number and temperature). The energetic distance between them, the effective ionization energy, decreases. At a certain density, it becomes zero, i.e., the band edge ``overtakes'' the exciton level which then merges with the continuum states in the band. The latter vanishing of the level (i.e., the breakup of the bound state) is usually referred to as Mott effect occurring at the Mott density. Note that this picture is strongly simplified and, in particular, does not include spectral broadening effects \cite{semkat2010,manzke2012}. For an in-depth discussion of the Mott effect see, e.g., \cite{semkat2009,manzke2012}.
As a consequence of the Mott effect, one observes the vanishing of the exciton absorption line experimentally \cite{heckoetter2018}.
Note that, after the crossing of the band edge and exciton level curves, the latter ones, of course, lose their meaning as discrete energy levels.

A further consequence from the Mott effect is a, for low temperatures, nearly steplike increase of the degree of ionization of the plasma from zero (only bound states) to one (fully ionized) \cite{rotesbuch}, known as Mott transition. The density at which this transition occurs is often referred to as Mott density, too, however, both ``Mott densities'' do not necessarily coincide (cf., e.g., Ref.\ \cite{semkat2009}).

Comparing $1S$- and $2P$-states (Fig.\ \ref{fig:mott1s2p}, upper and lower panel), we recognize the effect of the different signs of the line shifts: The $1S$-level shifts towards the band edge, while the $2P$-level ``runs away'' from the band edge. Note that the $2S$- and $2P$-level shifts differ only weakly, therefore, only the latter one is shown. The magnitude of the level shifts depends strongly on the Mott density and, thus, on the applied band edge model.

While the shift of the exciton level is given by the full dynamic solution for $\Delta E_{nl}^{(1)}$, Eqs.\ (\ref{DeltaE})--(\ref{imepsilonrpa}), on the one hand and its Debye limit [see Eq.\ (\ref{deltaedebye})] on the other hand, several approximations for the band edge are compared. Here, the ``simpler'' models (i) Debye shift $E_{\rm be}^{\rm Debye}$, Eq.\ (\ref{debye}), and (ii) $E_{\rm be}^{(0)}$, Eq.\ (\ref{bandkante}) with (\ref{bandkanters}) on the one hand and (iii) the more elaborated approximations $E_{\rm be}^{\rm iter}$ or $E_{\rm be}^{\rm mod}$, Eq.\ (\ref{bandkante}) with (\ref{bandkanteiter}) or (\ref{bandkanteiter1}), respectively, on the other hand differ largely in magnitude while the qualitative behavior is the same, cf.\ both panels of Fig.\ \ref{fig:mott1s2p}. Obviously, in these cases, the Debye approximation produces the strongest shift and, therefore, the lowest Mott density. The two iterative models deviate on the shown density scale only weakly from each other and give a significantly higher Mott density, about one and a half orders of magnitude.

The overestimation of the effective ionization energy reduction by the Debye approximation is well known, see, e.g., \cite{seidel95,arndt96,semkat2009}. However, in those previous works only the ground state or some of the lowest excited states have been considered.
\begin{figure}[tbh]%
\begin{center}
\includegraphics[width=0.9\columnwidth]{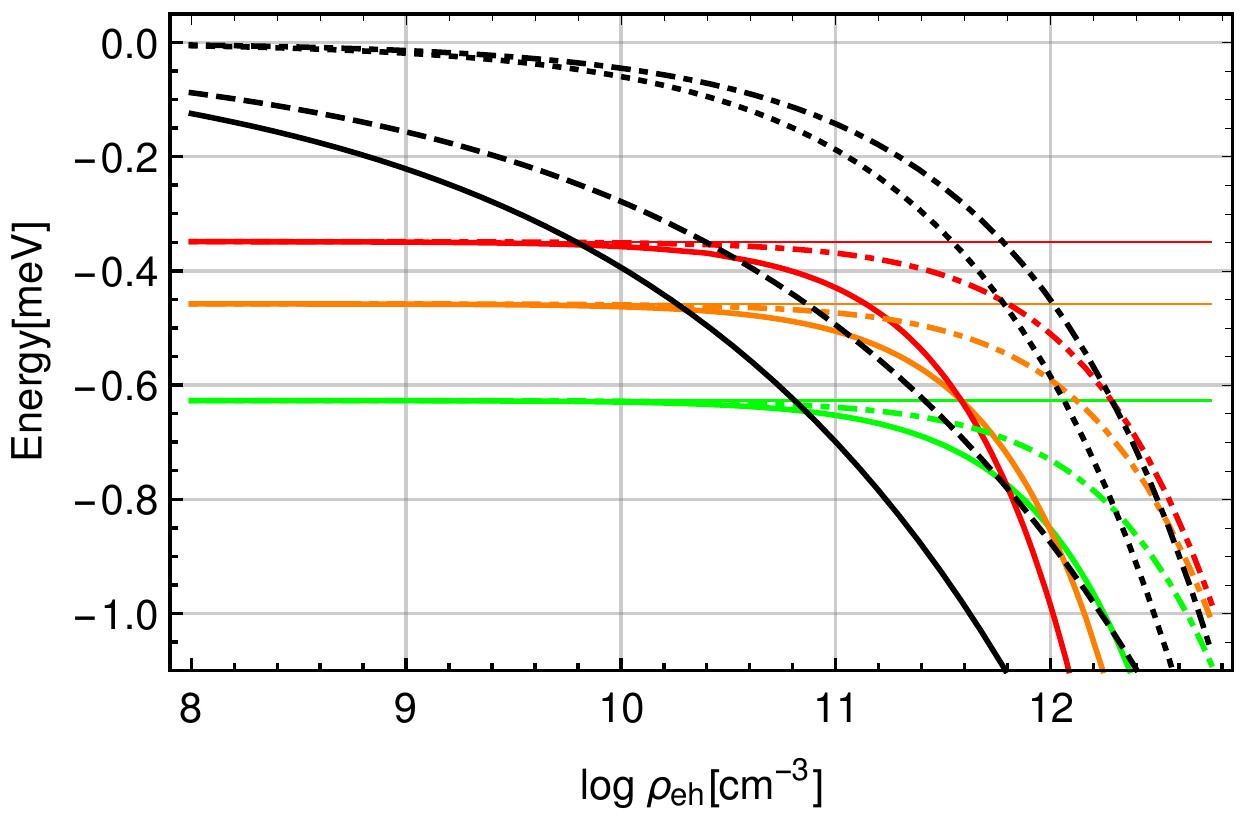}
\end{center}
\caption{Same presentation as Fig.\ \ref{fig:mott1s2p}, but for several $P$-exciton levels [$n=12$ (green lines), 14 (orange) and 16 (red)] for a carrier temperature of $T=10$ K.
}
\label{fig:mott121416}
\end{figure}

\begin{figure}[tbh]%
\begin{center}
\includegraphics[width=0.85\columnwidth]{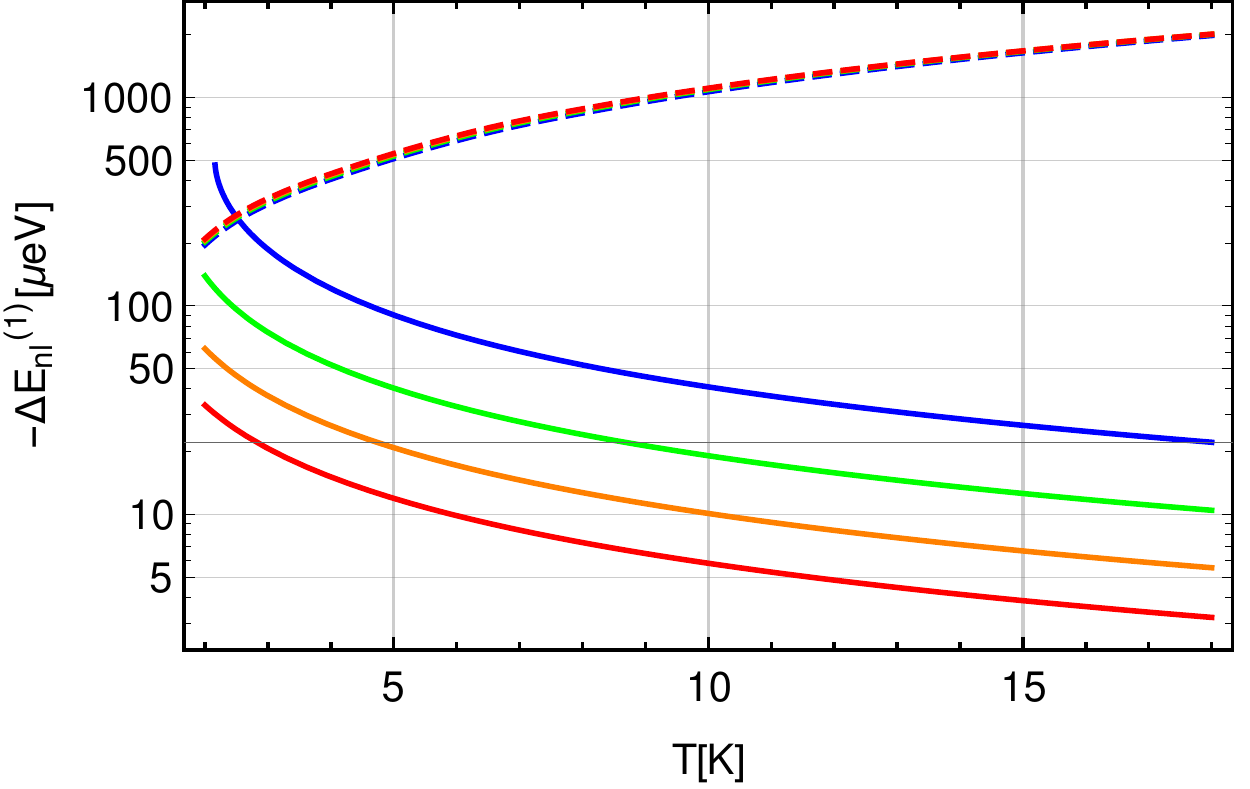}
\end{center}
\caption{Logarithmic plot of the maximum excitonic line shifts at the respective Mott density for the $n=10$ (blue lines), $n=12$ (green), $n=14$ (orange), and $n=16$ (red) $P$-excitons \textit{vs.}\ temperature $T$ using two different approximations: iterative solution of Eq.\ (\ref{bandkante}) with (\ref{bandkanteiter1}) for the band edge and fully dynamic line shift (solid) and Debye approximation for both quantities (dashed).
}
\label{fig:deltaemax10-16}
\end{figure}
A closer look onto a smaller energetic scale just below the band edge reveals a different sequence of the approximations for the band edge, see Fig.\ \ref{fig:mott121416}. There the iterative solutions decrease faster while the Mott effect for the high lying exciton levels occurs at much higher densities using the Debye and rigid shift models. Here, $n=12,14,16$ are chosen as examples. Another remarkable effect is that, on this scale of energies, both iterative solutions for the band edge, $E_{\rm be}^{\rm iter}$ and $E_{\rm be}^{\rm mod}$, differ quite considerably leading, e.g., for $n=12$ to Mott densities which differ by about an order of magnitude. Figure \ref{fig:mott121416} shows moreover that, although the line shifts at a certain density increase with $n$ (cf.\ Fig.\ \ref{fig:deltae10-16}), the maximum shifts before disappearing of the lines decrease with $n$ due to the decreasing Mott density, as shown in detail in Fig.\ \ref{fig:deltaemax10-16} for $E_{\rm be}^{\rm mod}$. The Debye approximation causes the opposite trend for the maximum shifts which are, in that case, nearly independent on $n$.

Note that the crossing of energy levels to be seen in Fig.\ \ref{fig:mott121416} (solid colored lines) is only an apparent effect since it occurs for densities higher than the respective Mott density where the excitonic states have already vanished in the band.

\begin{figure}[tbh]%
\begin{center}
\includegraphics[width=.85\columnwidth]{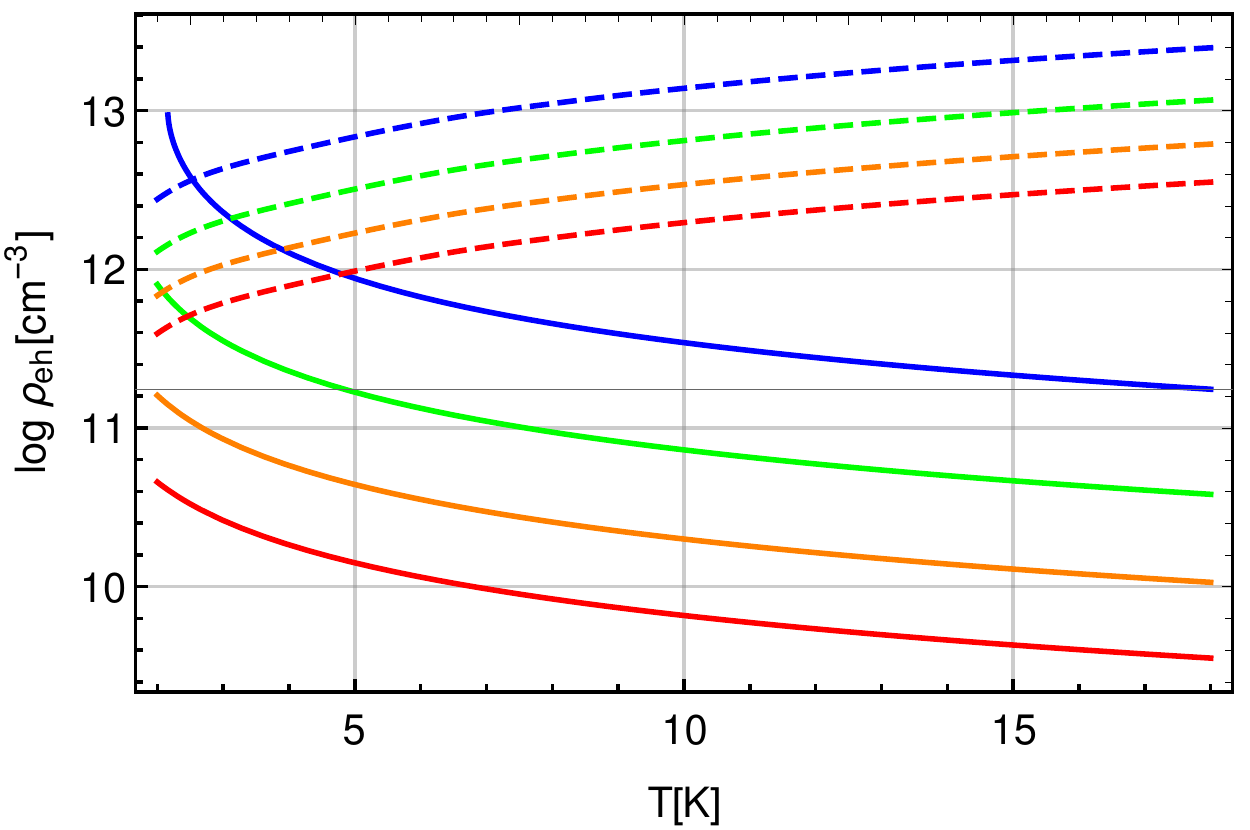}
\end{center}
\caption{Mott density \textit{vs.}\ temperature for several $P$-exciton levels. Line styles see Fig.\ \ref{fig:deltaemax10-16}.
}
\label{fig:mottdichte}
\end{figure}
The qualitatively different behavior of the iteratively determined band edge [we stick to the solution determined by (\ref{bandkanteiter1})] becomes apparent once more when looking at the resulting Mott densities $n_{\rm Mott}$ vs. temperature $T$ for several $n$, Fig.\ \ref{fig:mottdichte}. While the Mott density increases with increasing $T$ in the Debye case, the iterated band edge causes a decrease of $n_{\rm Mott}$.

\begin{figure}[tbh]%
\begin{center}
\includegraphics[width=.85\columnwidth]{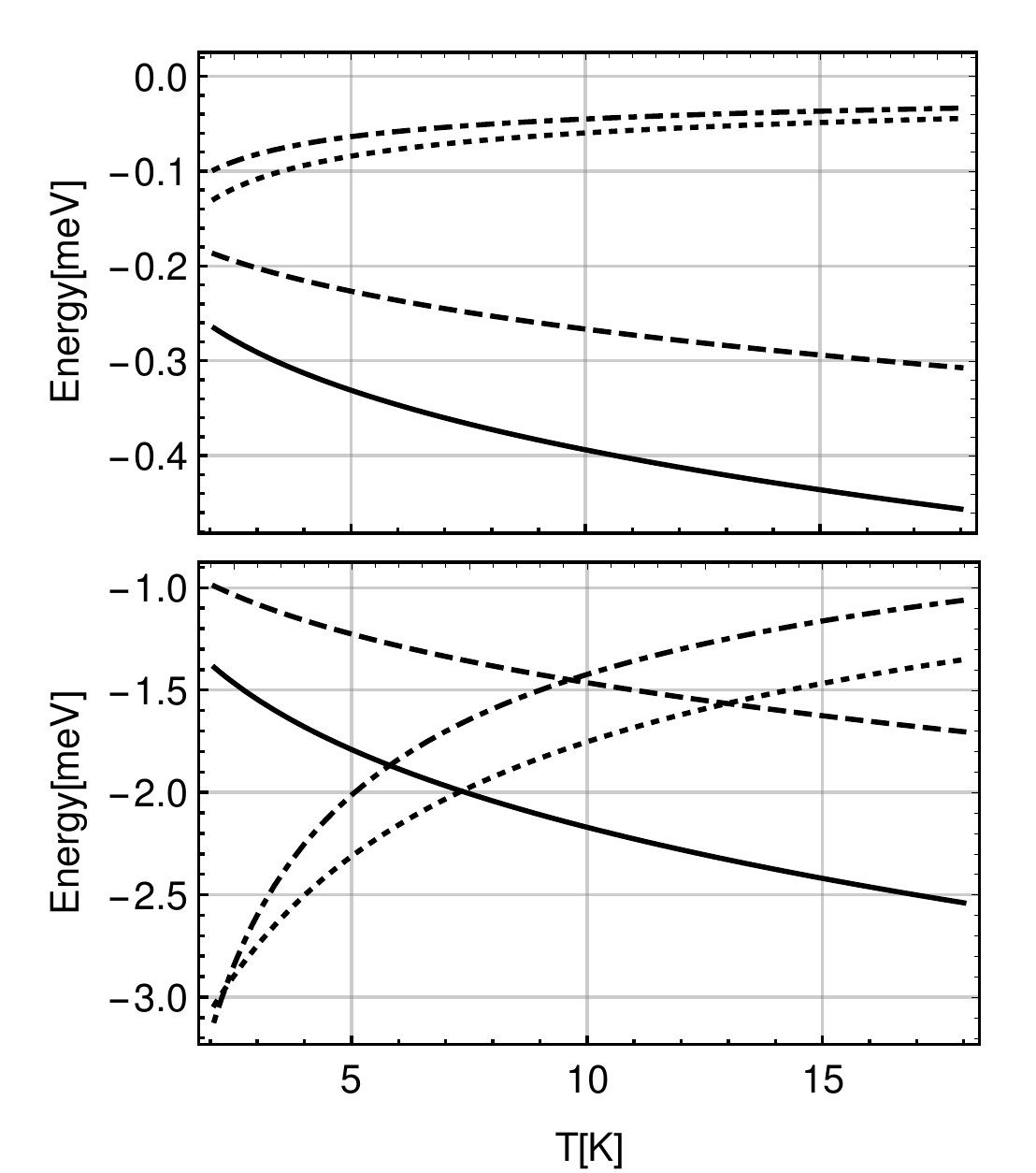}\\
\end{center}
\caption{Band edge \textit{vs.}\ temperature $T$ for a carrier density of $\rho_{\rm eh}=10^{10}$ cm$^{-3}$ (upper panel) and $\rho_{\rm eh}=10^{13}$ cm$^{-3}$ (lower panel): $E_{\rm be}^{\rm mod}$ (Eq.\ (\ref{bandkante}) with (\ref{bandkanteiter1}); solid line), $E_{\rm be}^{\rm iter}$ (Eq.\ (\ref{bandkante}) with (\ref{bandkanteiter}); dashed), $E_{\rm be}^{(0)}$ (Eq.\ (\ref{bandkante}) with (\ref{bandkanters}); dotted), and $E_{\rm be}^{\rm Debye}$ (Eq.\ (\ref{debye}); dash-dotted).
}
\label{fig:econt-T}
\end{figure}
In Fig.\ \ref{fig:econt-T}, the different band-edge shifts are depicted in dependence on temperature for two chosen densities. Debye shift as well as the explicit solution (\ref{bandkanters}) (note that the latter one is nothing else than the zeroth iteration step of Eq.\ (\ref{bandkante}) with (\ref{bandkanteiter}) or (\ref{bandkanteiter1}), respectively) decrease with increasing temperature. In contrast, the iterative, i.e., self-consistent, solution causes the absolute value of the shift to \textit{increase} with increasing temperature.
The analysis of density (cf.\ Figs.\ \ref{fig:mott1s2p} and \ref{fig:mott121416}) and temperature dependence of the band-edge shift reveals a scaling law the leading term of which is given by $E_{\rm be}^{\rm mod}=c\,\rho_{\rm eh}^{1/4}T^{1/4}$ with $c=6.92\cdot 10^2$ \textmu eV (\textmu m$^3$/K)$^{1/4}$ for low densities $\rho_{\rm eh}$. This is in contrast to the Debye shift which scales like $E_{\rm be}^{\rm Debye}\propto \rho_{\rm eh}^{1/2}T^{-1/2}$.

\begin{figure}[tbh]%
\begin{center}
\includegraphics*[width=0.98\columnwidth]{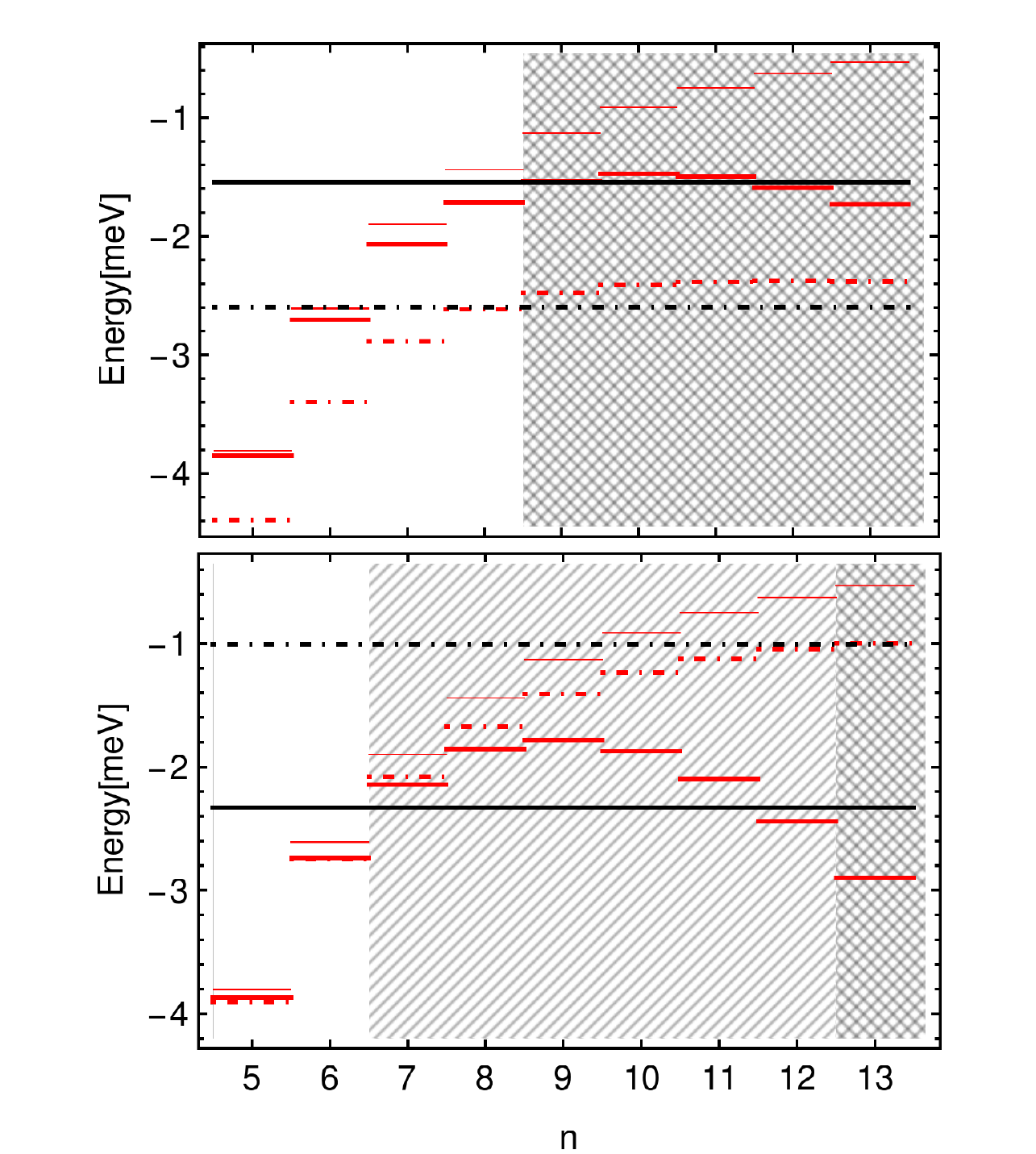}
\caption{Band edge (black lines) and exciton levels (red lines) \textit{vs.}\ principal quantum number $n$ for a carrier density of $\rho_{\rm eh}=10^{13}$ cm$^{-3}$ and two temperatures: $T=3$ K (upper panel) and $T=20$ K (lower panel). Line styles as in Fig.\ \ref{fig:econt-T}, i.e., solid lines: self-consistent model, dash-dotted lines: Debye model. Thin red lines denote the unperturbed exciton levels. Excitons with $n$ lying in the shaded areas cannot exist in both models (double-shaded) or exist only in the Debye model (single-shaded). Note that $n_{\rm max}=8$ for both models (upper panel) by chance.
}
\label{fig:mott3}
\end{center}
\end{figure}

The importance of assessing the different band edge approximations is pronounced again by Fig.\ \ref{fig:mott3}. It shows band edge and exciton levels of different principal quantum numbers $n$ for a given carrier density and two different temperatures. The highest observable exciton state (quantum number $n_{\rm max}$) in a measurement obviously depends very sensitively on the band edge model.
While $n_{\rm max}$ decreases for the self-consistent model (in the shown case from $n_{\rm max}=8$ at $T=2$ K to 6 at 20 K), it increases strongly for the Debye model (here from $n_{\rm max}=8$ at $T=2$ K to 12 at 20 K).

\section{Conclusions and outlook}

We have analyzed excitonic line shifts and band edge renormalization with special focus on higher lying (Rydberg) $P$-excitons in cuprous oxide. We applied a well-established many-body theoretical approach based on the theory of real-time Green's functions for this system which gained a lot of interest during the last few years. The excitonic line shifts have been calculated in first order perturbation theory with respect to the perturbation of the bound electron-hole state by many-body effects induced by the surrounding free carriers (electron-hole plasma). They are, except for the lowest principal quantum numbers, dominated by dynamical screening effects and, for the $P$-states, always negative, i.e., cause always a red shift of the lines.

For the band-edge shift, we have compared several approximations. One of them, a self-consistent, iterative solution, relies on a proposal by Seidel \textit{et al.} \cite{seidel95} which has been corrected here in order to treat band edge and quasiparticle energies on the same level of approximation. This self-consistently determined band-edge shift behaves qualitatively different compared to more conventional approximations like the Debye shift. In particular, the temperature dependence is just reverse to each other.

The vanishing of the exciton lines (i.e., the Mott effect) is, of course, determined by the interplay of line shift and band-edge shift. Here, the dependence of the Mott density on temperature and principal quantum number is qualitatively different for the different band edge models, too.

The ultimate test of the investigated models can only consist in comparing to experimental data. There are recent and ongoing experiments in the group of M. Bayer (Dortmund) measuring the absorption spectrum in various pump-probe scenarios, see e.g., \cite{heckoetter2018}. From these spectra, quantities like line and band-edge shifts as well as the vanishing of the excitonic lines in the band absorption are determinable. Our theory not only confirms that already very low electron-hole densities cause measurable effects on the spectrum, but is also able to predict qualitative trends (band-edge shift \textit{increasing} and maximum achievable principal quantum number \textit{decreasing} with increasing temperature) in contradiction to the widely used semiclassical Debye approach.

Our results confirm on the one hand the importance of investigating the action of an electron-hole plasma on Rydberg exciton states in order to get a deeper understanding of their properties. On the other hand, they show the great potential of Rydberg excitons both experimentally---as a kind of sensor for the detection of very low carrier concentrations---as well as theoretically by learning from their behavior to verify or falsify certain many-body theoretical models.

An in-depth discussion of the recent experiments and their results as well as a detailed comparison to the theoretical predictions will be done in a forthcoming paper \cite{grossespaper}.

\acknowledgments

We wish to thank S. O. Kr\"uger, W.-D. Kraeft, R. Schwartz, Th.\ Bornath, and G. R\"opke (Rostock), M. Bayer (Dortmund), and A. Alvermann (Greifswald) for many fruitful discussions and K. Sperlich (Rostock) for valuable help with the numerics.
D.\ S.\ gratefully acknowledges support by the Deutsche Forschungsgemeinschaft (project number SE 2885/1-1).


\begin{thebibliography}{20}

\bibitem{nature2014} T. Kazimierczuk, D. Fr\"ohlich, S. Scheel, H. Stolz, and M. Bayer, Nature \textbf{514}, 343 (2014).

\bibitem{thewes2015} J. Thewes, J. Heck\"otter, T. Kazimierczuk, M. A{\ss}mann, D. Fr\"ohlich, M. Bayer, M. A. Semina, and M. M. Glazov, Phys. Rev. Lett. \textbf{115}, 027402 (2015).

\bibitem{feldmaierzielinska1617} M. Feldmaier, J. Main, F. Schweiner, H. Cartarius, and G. Wunner, J. Phys. B: At. Mol. Opt. \textbf{49}, 144002 (2016); S. Zieli\'{n}ska-Raczy\'{n}ska, D. Ziemkiewicz, and G. Czajkowski, Phys. Rev. B \textbf{94}, 045205 (2016); S. Zieli\'{n}ska-Raczy\'{n}ska, D. Ziemkiewicz, and G. Czajkowski, Phys. Rev. B \textbf{95}, 075204 (2017).

\bibitem{assmannostrovskaya2016} M. A{\ss}mann, J. Thewes, D. Fr\"ohlich, and M. Bayer, Nature Materials \textbf{15}, 741 (2016); E. A. Ostrovskaya and F. Nori, Nature Materials \textbf{15}, 702 (2016).

\bibitem{walther2018a} V. Walther, S. O. Kr\"uger, S. Scheel, and Th. Pohl, Phys. Rev. B \textbf{98}, 165201 (2018).

\bibitem{walther2018b} V. Walther, R. Johne, and Th. Pohl, Nature Commun. \textbf{9}, 1309 (2018).

\bibitem{heckoetter2018} J. Heck\"otter, M. Freitag, D. Fr\"ohlich, M. A{\ss}mann, M. Bayer, P. Gr\"unwald, F. Sch\"one, D. Semkat, H. Stolz, and S. Scheel, Phys. Rev. Lett. \textbf{121}, 097401 (2018).

\bibitem{grossespaper} R. Schwartz \textit{et al.}, to be published.

\bibitem{rotesbuch} W. Ebeling, W. D. Kraeft, and D. Kremp, \textit{Theory of Bound States and Ionization Equilibrium in Plasmas and Solids},
Akademie--Verlag, Berlin (1976); Mir, Moscow (1979).

\bibitem{gruenesbuch} W. D. Kraeft, D. Kremp, W. Ebeling, and G. R\"opke, \textit{Quantum Statistics of Charged Particle Systems}, Akademie--Verlag, Berlin and Plenum Press, London (1986).

\bibitem{zimmermann88} R. Zimmermann, \textit{Many-Particle Theory of Highly Excited Semiconductors}, Teubner--Texte zur Physik Bd. 18, BSB B. G. Teubner Verlagsgesellschaft, Leipzig (1988).

\bibitem{kremp2005} D. Kremp, M. Schlanges, and W.-D. Kraeft, \textit{Quantum Statistics of Nonideal Plasmas}, Springer Series on Atomic, Optical, and Plasma Physics 25, Springer--Verlag, Berlin, Heidelberg (2005).

\bibitem{seidel95} J. Seidel, S. Arndt, and W. D. Kraeft, Phys. Rev. E \textbf{52}, 5387 (1995).

\bibitem{footnote0} Note that there are other possibilities to define the band edge \cite{kraeft2019}.

\bibitem{kraeft2019} W.-D. Kraeft, private communication.

\bibitem{footnote1} The method is described in Ref.\ \cite{seidel95}, see Eqs.\ (31) and (32) therein.

\bibitem{5maenner} R. Zimmermann, K. Kilimann, W. D. Kraeft, D. Kremp, and G. R\"opke, Phys. Status Solidi B \textbf{90}, 175 (1978).

\bibitem{fehr94} R. Fehr and W. D. Kraeft, Phys. Rev. E \textbf{50}, 463 (1994).

\bibitem{footnote2} Note that there is a sign error in the imaginary part of $\varepsilon$ in \cite{fehr94}, Eq.\ (2.8).

\bibitem{arndt96} S. Arndt, W. D. Kraeft, and J. Seidel, Phys. Stat. Sol. B \textbf{194}, 601 (1996).

\bibitem{stolz2018} H. Stolz, F. Sch\"one, and D. Semkat, New J. Phys. \textbf{20}, 023019 (2018).

\bibitem{haug78} H. Haug and D. B. Tran Thoai, Phys. Status Solidi B \textbf{85}, 561 (1978).

\bibitem{semkat2010} D. Semkat, F. Richter, D. Kremp, G. Manzke, W.-D. Kraeft, and K. Henneberger, J. Phys.: Conf. Ser. \textbf{220}, 012005 (2010).

\bibitem{manzke2012} G. Manzke, D. Semkat, and H. Stolz, New J. Phys. \textbf{14}, 095002 (2012).

\bibitem{semkat2009} D. Semkat, F. Richter, D. Kremp, G. Manzke, W.-D. Kraeft, and K. Henneberger, Phys. Rev. B \textbf{80}, 155201 (2009).

\end{thebibliography}
\end{document}